# A comprehensive review of Chronic Kidney Disease of Unknown Etiology

Adoni Fernando[1], Nivethika Sivakumaran[1*]
[1](*Undergraduate Department of Biomedical Science, International College of Business and Technology, No. 36, De Kretser Place, Bambalapitiya, Sri Lanka*)

**ABSTRACT :**
*This review was written to provide a comprehensive summary of the suggested etiologies of Chronic Kidney Disease of Unknown Etiology (CKDu) in Sri Lanka. In this review, Chronic Kidney Disease (CKD) is explained in detail and its known etiologies are discussed. CKDu is defined and its epidemiology is discussed, with the compilation of statistic from over 15 research papers through the years 2000 to present.*

**ABBREVIATIONS:**
*CKDu: Chronic kidney disease of unknown etiology, CKD: Chronic Kidney Disease, CKDmfo: Chronic Kidney Disease of multifactorial origin, CINAC: Chronic Interstitial Nephritis in Agricultural Communities, NCP: North Central Province (of Sri Lanka), CRF: Chronic Renal Failure, CRD- Chronic Renal Disease, GFR- Glomerular Filtration Rate, NFK- National Kidney Foundation (USA), KDQOI- Kidney Disease Quality Outcome Initiative, ESRD- End Stage Renal Disease, BMI- Body Mass Index, NSAID- Non Steroidal Anti Inflammatory Drug, DDT- Dichlorodiphenyltrichloroethane, Cd- Cadmium, Pb- Lead, U- Uranium, N- Nitrogen, K- Potassium*

## I. INTRODUCTION

Chronic kidney disease (CKD) is defined as kidney damage or a decreased glomerular filtration rate. CKD is characterized by progressive destruction of renal mass, with irreversible sclerosis and loss of nephrons over a period of time, depending on the underlying etiology. Diabetes, hypertension and different forms of glomerular nephritis are known etiologies of CKD [1]. The rising prevalence of non-infectious diseases such as diabetes mellitus and hypertension suggests that CKD incidence will continue to rise [2]. Over the last two decades, the epidemic of a new strain of CKD of unknown etiology in rural farming communities of Sri Lanka, has elevated the concern of the disease. This new strain has been observed to be endemic in agricultural and rural regions. CKDu may also be referred to as CKD of multifactorial origin (CKDmfo), chronic agrochemical nephrology (CAN) or chronic interstitial nephritis of agricultural communities (CINAC) [3]. The disease is prevalent especially in the North Central Province (NCP). It is observed in mainly the dry regions that may utilize irrigation methods for farming. It disproportionately affects males of poor socio-economic background and workers of paddy farming [4, 5]. In Sri Lanka there is an estimated affected population of 400,000 people and death toll of around 20,000 people. CKDu is not exclusive to Sri Lanka, and many cases of CKD of unknown etiology have been reported in different parts of the world; e.g., Costa Rica [6], India [7], Egypt [8], Nicaragua [9], and more. However, in Sri Lanka, there is limited knowledge on the exact prevalence and geographical distribution of CKDu at this time.

Even with the suggestion of many etiologies, the limited supporting evidence means that the exact cause of the disease is still unknown. Water contamination is a popular suggestion, but the evidence supporting the claim is insufficient. The role of environmental exposure to high concentrations of nephrotoxicants (such as, Cd, Pb, and U), which may accumulate and cause functional and structural damage in renal tissue, is a weighty hypothesis [10]. Farming and the use of fertilizer or other agrochemicals are also considered risk factors [11, 12]. High levels of as exposure is also suggested to be an etiological factor in CKDu [13].

Water used for consumption is thought to be a major source of contaminants that lead to CKD. The contaminants may seep into water from the natural environment (geological sources), but the intensive use of agrochemicals to increase the production of agricultural resources is also a significant possibility of water contamination. The local water bodies may be polluted with seepage from irrigation or farming, especially those that lay close to fields used for cultivation. A systematic assessment of water quality is therefore needed, in order to understand risk related to exposure and assess the potential health effects [14].

This review attempts to provide a comprehensive understanding of CKDu and its prevalence in Sri Lanka. It attempts to document the possible etiologies of CKDu, put forward by research throughout the year 1997 to present.

The endemic regions in North Western Province of Sri Lanka is shown in Fig.1. Polpithigama is identified as a high risk geographic area for CKDu (Screening Guidelines Chronic Kidney Disease Sri Lanka 2017 Epidemiology Unit Ministry of Health, 2017). Nikawewa has a high prevalence of the disease and has two specialized nephrology units, catering to CKD patients.

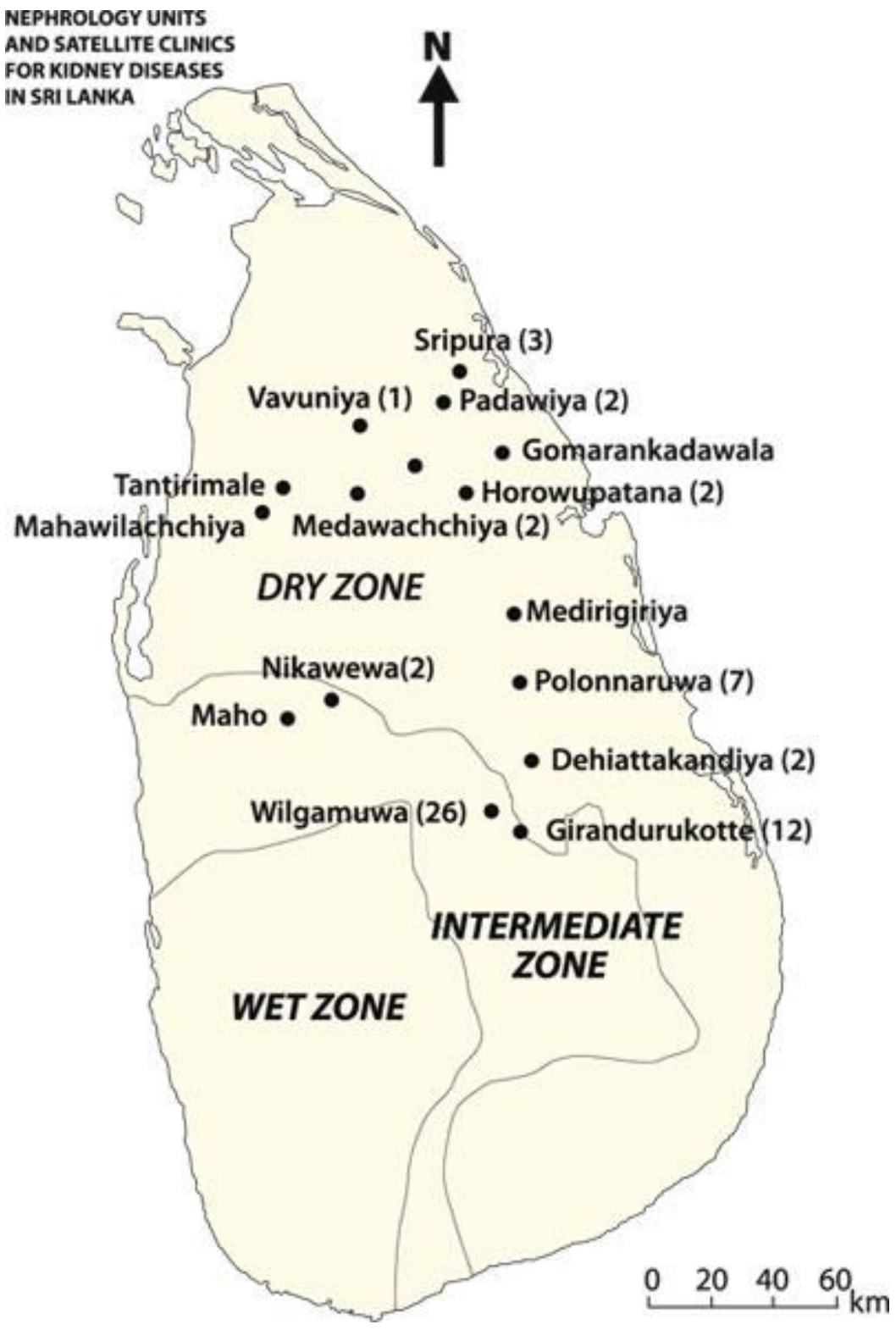

*Figure 1. Nephrology units and satellite clinics for kidney diseases in Sri Lanka* (Daily News, 2017)

## II. CHRONIC KIDNEY DISEASE

Chronic Kidney Disease is recognized as a global public health problem. The disease is coined from the kidneys functions progressively diminishing over time due to the progressive damage of the renal mass. CKD is also known as Chronic Renal Failure (CRF), Chronic Renal Disease (CRD) and Chronic Kidney Failure (CKF). CKD can be defined as structural or functional anomalies of the kidneys or a decreased glomerular filtration rate (GFR), persisting for a period of three months, (<60 ml/min or 1.73 m2).

2.1. Functions of the kidneys

The main functions of the kidneys are; filtration of waste products and excess water from blood to be excreted through urine, regulation of blood pressure, production of erythropoietin (important for formation of red blood cells, and maintaining balance of salts and chemicals in bloodstream at the right level(NFK, www.kidney.org). CKD is a slowly progressing disease, and often goes unnoticed and underdiagnosed. It is usually asymptomatic until histopathological features such as tubulointestitial fibrosis and tubular atrophy, observed in renal biopsies. Usually when the signs and symptoms manifest, the disease has reached to a severity, for which the cure is a kidney transplant, without which there is a 100% mortality rate. The conditions of a CKD patient irreversibly worsen over time.

2.2. Stages of CKD

CKD is diagnosed based on definition of the Kidney Disease Quality Outcome Initiative (KDQOI) of the National Kidney Foundation (NFK). The functional damage of kidneys can be assessed by GFR, and the disease can be classified into 5 stages such as, stage 1: Kidney damage with normal or increased GFR (>90 mL/min /1.73 m2), stage 2: Mild reduction in GFR (60-89 mL/min/1.73 m2), stage 3: Moderate reduction in GFR (30-59 mL/min/1.73 m2), stage 4: Severe reduction in GFR (15-29 mL/min/1.73 m2), and stage 5: Kidney failure (GFR < 15 mL/min/1.73 m2 or dialysis).

CKD is usually detected when the kidneys function drops to 25% of normal kidney function, whereas symptoms appear when the function plummets to 1/10th of the normal level. Each kidney has approximately one million nephrons, the functional unit of kidney, which filter wastes from the body. The human body can maintain the normal GFR, even when there is a gradual destruction of the nephrons. This happens by mechanisms to compensate for the loss, such as hypertrophy of renal tissue and hyper filtration by the surviving nephrons. This occurs regardless of the underlying etiology of the CKD, and allows blood to be filtered and the plasma solutes to be cleared as it would in normal conditions. When GFR decreases to 50% of normal, the level of plasma solutes such as creatinine and urea show a significant increase. The creatinine in plasma will be approximately double that of normal, when GFR drops to 50%; i.e., when the plasma creatinine reaches 1.2 mg/dL from the baseline value of 0.6 mg/dL, it represents halving of the functional nephron mass, although it is still within the defined reference range.

2.3. Known etiologies of Chronic Kidney Disease

Certain common chronic diseases can lead to chronic kidney disease. These diseases can damage the renal tissue permanently causing damage and giving rise to CKD.

2.3.1. Diabetes mellitus

CKD was thought to develop as a result of diabetes and was known as "diabetic nephropathy". The elevated levels of blood glucose leads to an increased strain on the nephrons of the kidneys when filtering blood. It is estimated that moderate to severe CKD is found in 15-23% of patients that have a history of diabetes [15].

2.3.2. Hypertension

High blood pressure is a leading cause of CKD globally. Hypertension damages the blood vessels and may lead to a reduced blood flow to kidneys. The elevated pressure can also damage the nephrons assisting renal damage. More than 50% of the CKD patients suffer from hypertension (High Blood Pressure and Chronic Kidney Disease www.kidney.org National Kidney Foundation’s Kidney disease outcomes Quality initiative, n.d.).

Diabetes mellitus and hypertension are the identified major etiologies of CKD.

In addition, Tubulointerstitial disease has been seen to have a strong link to CKD in Sri Lanka. There has been a significant prevalence of CKD observed in toxic nephropathies due to tubulointerstitial disease. There are a few identified causes of tubulointerstitial disease such as heavy metals, radiation nephritis, chronic hyperkalemia, chronic hypercalcemia, etc. [16]. Progressive kidney diseases that lead to CKD may also be caused by proteinuria, systemic hypertension, elevated nephrotoxins or decreased perfusion, smoking, uncontrolled diabetes, illegal drug abuse, physical injury, and many more.

## III. CHRONIC KIDNEY DISEASE OF UNKNOWN ETIOLOGY (CKDu)

The Health Ministry Circular defines CKDu as renal damage with no past history or current treatment for diabetes mellitus, chronic and/or severe hypertension, snake bites, urological disease of known etiology or glomerulonephritis; with normal HbA1C (glycated Hb<6.5%); and blood pressure at <160/100 mmHg or 140/90 mmHg on up to two antihypertensive drugs. CKDu endemic occurrences was noted recognized in 1990, and has been increasing to date.

### 3.1. Epidemiology of Chronic Kidney Disease of Unknown Etiology (CKDu)

The outbreak and prevalence of CKD of unknown etiology was first recognized in the early 1990's in the NCP, namely Dehiattakandiya, Padaviya, Girandurukotte, Kabithigollawa, Medawachiya, Medirigiriya and Nikawewa of Sri Lanka, according to hospital records as shown in Fig.2.It was later found that the disease was endemic to agricultural communities.

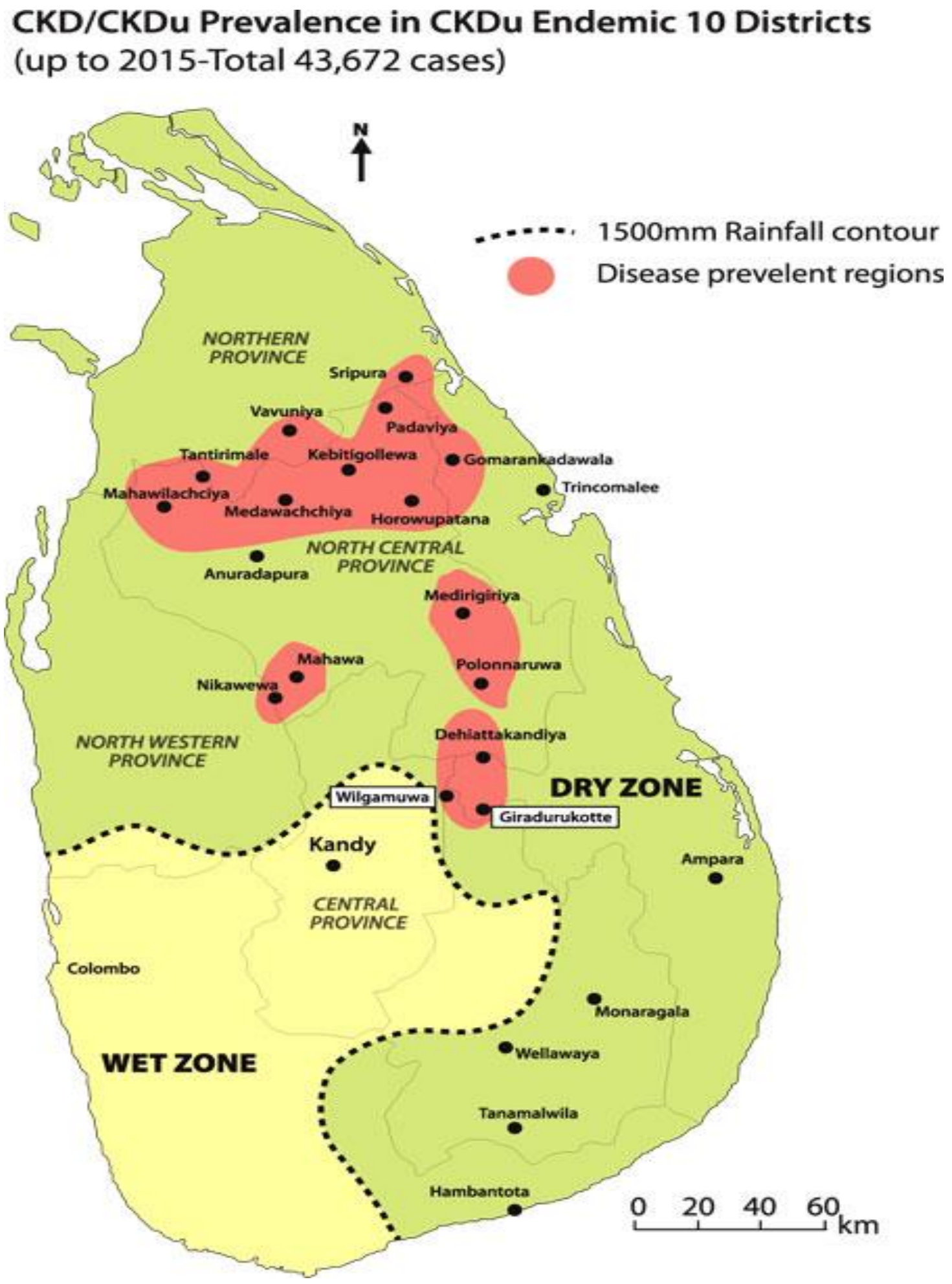

Figure 2. CKD/CKDu prevalence (2015) (Daily News, 2017)

In February 2017, the Epidemiology Unit of the Ministry of Health, Nutrition & Indigenous Medicine released an epidemiology report publication (Vol. 44 No. 07), highlighting the regions in which CKDu was endemic and had a very high prevalence (Screening Guidelines Chronic Kidney Disease Sri Lanka 2017 Epidemiology Unit Ministry of Health, 2017).

The identified high risk geographic areas for CKDu in Sri Lanka are shown in Table 1.

Table 1. High risk geographic areas for CKDu in Sri Lanka

| Province | District | DS Division |
|---|---|---|
| North Central | Anuradhapura<br>Polonnaruwa | All<br>All |
| North Western | Kurunegala | Polpithigama&Giribawa |
| Eastern | Ampara<br>Trincomalee | Dehiattakandiya<br>PadaviSripura |
| Uva | Badulla | Mahiyanganaya&Rideemaliyadda |
| North | Mullaitivu<br>Vavuniya | Welioya<br>Vavuniya&Vavuniya South |
| Central | Matale | Wilgamuwa |

The disease was seen to be prevalent in regions around water reservoirs that are used for irrigation purposes, and mostly among farmers [16]. A population screening of endemic regions shows that a cohort of the younger members of the population is also affected by the disease, with a distinctly high incidence reported in Medawachchiya. The prevalence of CKDu in the overall population was measured to be 3.7% (n=4107), whilst the prevalence of disease in the adult (>18 years of age) population was noted to be 5% (n=2600). The ratio of distribution of disease in respect to gender (female: male) was 1:1.3 [17].

According the WHO report on the "Kidney Disease of Uncertain Etiology (CKDu) in Sri Lanka", there is a 16.9% age standardized prevalence of CKDu among females within age group 15-70 years, and a 12.9% prevalence of males in the same age group. The main sources of water used for consumption, in CKDu affected regions, are shallow and deep wells. It is reported that in the areas with very high prevalence of CKDu, nearly 87% of the population utilize either dug well and tube well water for their daily needs. Water is used in cooking and is usually boiled before consumption, but is otherwise untreated [18].

The number of patients treated for kidney disease have shown an increase in the years 2001-02, and the main identified cause of renal damage is CKDu. In another report by Athuraliya, it was found that 2-3% of the population are affected by CKDu, in identified endemic areas. It was also reported around 5000 patients undergoing treatment, in places with high prevalence of CKDu, due to renal failure at the time, and more than 9000 patients are followed up at the regional hospitals. Each year there are about 2000 patients diagnosed with the end stage of renal disease, with urgent need of kidney transplantation; without which, there is a 100% mortality rate [17].

Population screening carried out by a technique of multistage sampling, had resulted in a CKDu point prevalence of about 2-3% in subjects over 18 years of age [21]. Grounds view 2012 reported around 1500 deaths due to CKD nationwide. There was a reported 12.9% rate of CKDu increase [12].

A hospital based CKD registry was established in Sri Lanka by the WHO in 2009, with four hospitals in Anuradhapura, Medirigiriya, Polonnaruwa and Medawachchiya, in the NCP. There was a total of 1997 CKD patients, of which 775 cases (39%) were CKDu cases. CKDu as a percentage of the total number of cases of CKD is as follows: Anuradhapura- 48%, Medirigiriya- 25%, Pollonaruwa- 14%, Medawachchiya- 13%. 72% of the total patients of CKDu was seen to be male, whilst 28% was female. Another study carried out by the WHO found that 15% of the population of the NCP are affected by CKDu [19, 20].

CKDu was classified as a middle farmers disease, as the mean age of the male patients with CKDu was 45.5 years, whilst it was found to be 47.4 years in females [22]. However, in a previous study carried out in 2007, it was found that the mean age of male CKDu patients was 56.7 years, and the mean age of females was found to be 54.2 years [23]. This decreasing mean age of the CKDu patients supports that CKDu is a suggestive of the possible cumulating effects of toxins in the pathogenesis of the disease [24].

3.2. Suggested etiologies of Chronic Kidney Disease of Unknown Etiology (CKDu)

There are many suggested explanations for the etiology of the CKDu, the hypotheses that have the highest significance in CKDu in Sri Lanka have been explained below.

3.2.1. Cadmium

It was put forward that the high contents of cadmium being consumed by the population in the dry zone was the cause of the high prevalence of CKDu. It was reported that large amounts of cadmium entered the body through water, fish and lotus rhizomes [25]. The study said that the high content of cadmium could be due to the mass use of phosphate fertilizer contaminated with cadmium. However, this theory was refuted in another 2010 study, which found out that there were higher cadmium levels in regions in the non-endemic region whilst compared to the endemic regions. The clinical evidence of this study, also didn't show any other diseases that appear when there is high exposure to cadmium, such as renal calculi, respiratory effects and bone disease, which usually manifest prior to renal damage.

3.2.2. Arsenic

The presence of calcium arsenate crystals deposited in post-mortem test of CKD affected kidneys, suggested that consumption of arsenic and calcium in increased levels was the cause of CKDu. The areas that the study was carried out had an increased level of as and hardness of water. The calcium of the hard water, along with the arsenic leads to the formation of the calcium arsenate crystals that bind to arsenic transporters in the liver and moves to the renal tissue where it interrupts the kidneys antioxidant defense.

The study found 0.02-0.1 ppm of arsenic in drinking water, greater than that of the WHO recommended value of 0.01 ppm. The clinical assessment discovered a high level of arsenic in hair (mean of 460 ppb). Even though the Sri Lankan government has banned the use of arsenic in pesticides in 2001, there is speculation that it is still being used illegally. A study presented by Dr. Jayasumana, showed that there was 0.18-2.58 ppm of arsenic found in the agrochemicals used by Sri Lankan farmers. This study however brings to light, that these agrochemicals are used throughout Sri Lanka, but only certain regions show high prevalence of CKDu, whilst other regions remain non-endemic [24,26].

3.2.3. Aluminium Fluoride

According to one study, the CKDu endemic regions had fluoride rich water and the population relied on sub-standard utensils for cooking and storage of water. The acidic conditions during cooking, and the emanation of aluminum from these pots and pans during different levels of fluoride was studied. The amount of aluminum leached into the water in the absence of an acidic component was relatively small (1.20 ppm- after 10 mins of boiling a 6ppm fluoride solution). In acidic conditions of pH 3.02 (using tamarind), even in the absence of fluoride, there was 18 ppm aluminum leached from the utensils; with regular increases in the emanated aluminum with the increase in fluoride content. After 10 mins of boiling in a 6ppm fluoride solution, with acidic additives, the amount of aluminum that had seeped in to solution was 29 ppm. It was observed that the fluoride in water encourages aluminum leaching into the food being cooked, and the aluminofluoride complexes possibly can be an etiological factor causing CKDu.

The hypothesis suggested that since there is greater use of inferior quality aluminium utensils, that are made of recycled scrap metal and alloys, there is high chance that they are also contaminated with other heavy metals, such as lead, which are potentially poisonous to the renal tissues, contributing to CKDu [14, 27, 28].

However the study didn't explain how the CKDu is restricted to certain geographical regions, even though the use of these low quality utensils are present throughout the nation.

3.2.4. Hardness of water

A strong positive correlation ($p<0.008$) was seen between hard water and the arsenic content of water in CKDu endemic regions, when compared to the regions with less prevalence of CKDu, which showed a weak correlation. This study also implicated that the consumption of the hard water along with the elevated levels of arsenic was a possible etiological factor of CKDu [29].96% of the CKDu patients consumed hard water for a minimum of 5 years [22].

According to a preliminary study on the hard water of three districts of NCP Padaviya, Medawachchiya and Kabithigollawa and Ampara district of the WP. The study found high levels of calcium, magnesium and zinc in the water used for consumption. However there was no detectable traces of arsenic, lead or mercury. The water was tested from the regions with significantly high prevalence of CKD. The preliminary findings did include brittle bones, which is a symptom of hard water [19].

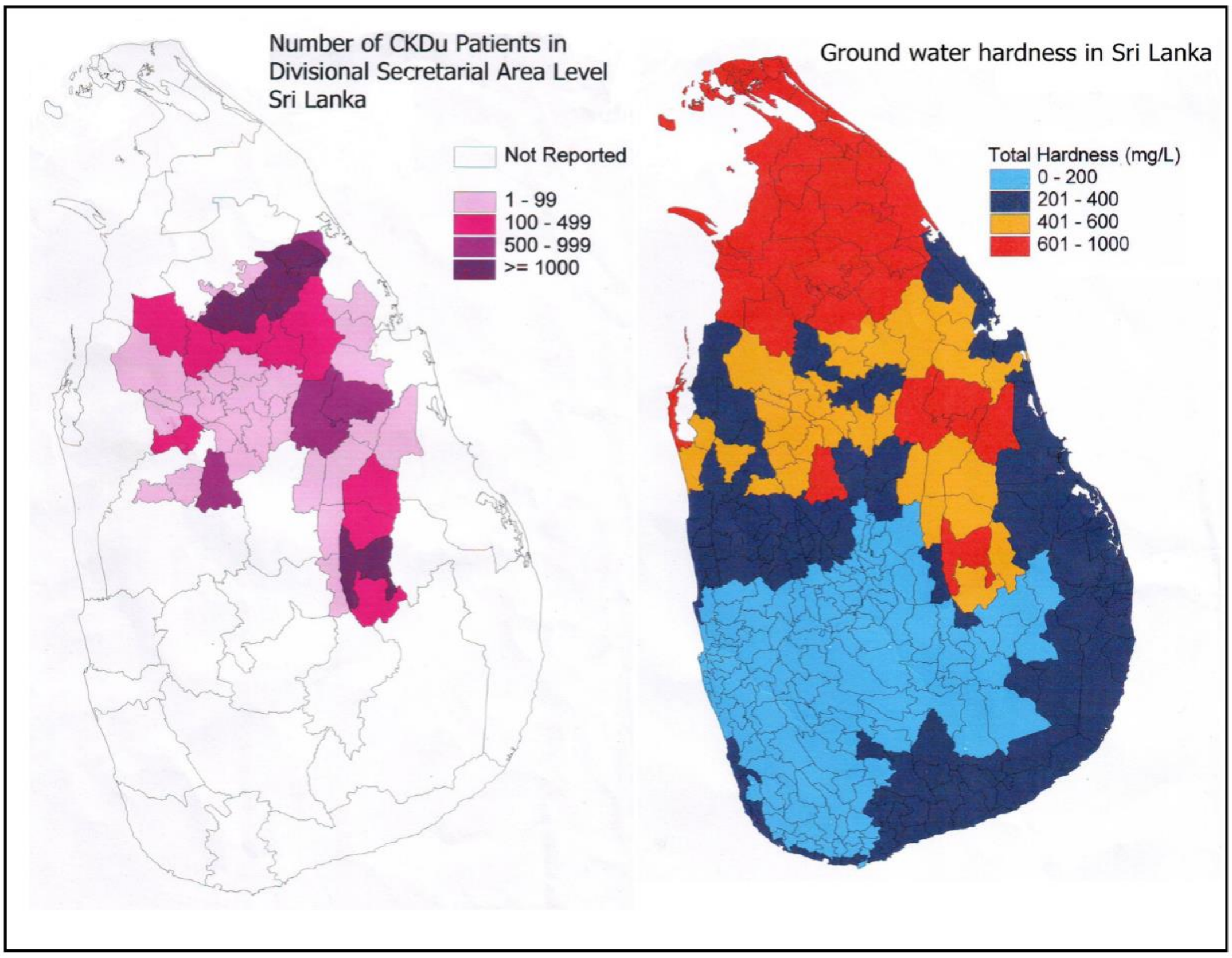

Figure 3-Distribution of hard water (left) and distribution and prevalence of CKDu in Sri Lanka (right)in Sri Lanka(Jayasumana, Gunatilake and Senanayake, 2014)

3.2.5. Low selenium

Selenium concentration of 80-95 µg/l is needed by the body and has been shown to protect the kidneys from oxidative stress. In the study carried out in 2013, it was seen that 38% of CKDu patients showed a serum selenium concentration below 80 µg/land below 90 µg/l in 63% of individuals affecting CKDu. Therefore it was suggested that low selenium levels may be a contributory factor making the kidneys more vulnerable to oxidative damage that can be caused by heavy metals and metalloids. Deficiency of selenium seen in individuals with CKD suggest that it is possibly a predisposition to the development of CKDu [12].

3.2.6. Pesticides

The popular belief among the locals of the affected regions is that the use of pesticides and its bioaccumulation, is the causative factor of the renal complications and CKD they face. It is believed that the pesticides have leached into the drinking water and the food grown using these agrochemicals. The local fish varieties have perished and effects are even seen in their cattle. The Minneriya Irrigation System caused many people from the wet zones of the country to settle into the NCP in the 1950's. The older generation (settlers) in this area are not as affected as the younger cohort (mostly children of the settlers) who began farming in the endemic regions. The condition worsened after the introduction of the green revolution [19]. The pesticide usage in paddy cultivation and vegetable growing regions is significantly high, though there is no provincial data available. These regions include those that have recorded significantly higher levels of CKDu, such as Polonnaruwa, Anuradhapura, Ampara, and Badulla.

3.2.7. Ayurvedic medicine

Aristolochic acid was suggested to be an etiological factor of BEN, as the wheat used in the Balkan regions was found to be contaminated with *Aristolochia clemantis.*

Therefore, this initiated studies to be carried out into the ayurvedic medicine used in Sri Lanka, to test for the presence of Aristocholic acid to explain renal disease. Research conducted have observed that the use of these herbal medication produces nephrotoxic effects in experimental animals and patients. It was observed that around 90% of the ayurvedic practitioners used alternative medical remedies, whilst only 10% used *Aristoclochiaindica* (Sapsanda) as an ingredient in their recipes. Sapsanda was used as a remedy for few diseases as well as for poisonous bites, the most common method of use was external application. Therefore, it was concluded that the use of nephrotoxic herbal medication was not a causal factor for CKDu found in the NCP [30].

3.2.8. Glyphosate

Glyphosate is the most widely used herbicide in the endemic areas of disease. Glyphosate and related compounds are known to have strong chelating properties, however, no serious consideration has been given to the possible effects of these compounds on human health. This could be due to the fact that glyphosate in its original form is known to be an easily degradable compound in the natural environment, and its effectiveness as an herbicide.

The GMA lattice hypothesis states that glyphosate can be absorbed through the dermis or through inhalation into body directly. It can also leach into the water bodies to form glyphosate metal complexes (the endemic regions are known to have elevated hardness). This allows another route for glyphosate to enter into body. These glyphosate compounds can lead to oxidative stress, nitrosative stress, apoptosis and necrosis of the renal tissue; leading to glomerular sclerosis, glomerular collapse, proximal tubular damage and tubulointerstitial damage. This evidenced by the elevated excretion of As, Cd, heavy metals, glyphosate and AMPA in urine of healthy individuals, whilst patients with CKDu have significantly lower levels of these compound excreted in urine. Although glyphosate alone might not be the cause of CKDu, it is arguable that the raised levels of glyphosate in endemic regions, coupled with localized geo environmental factors and nephrotoxic substances causes renal damage, which may lead to CKDu [29].

3.2.9. Bioaccumulation

Most CKDu endemic areas are around tanks and reservoirs. The big tanks are usually home to many plants and aquatic animals. The bioaccumulation of pesticide residues, toxins and heavy metals in these organisms is a major area of concern. Both the Mahaweli Irrigation system and the Minneriya Irrigation System both exist in the dry zone, namely NCP. Water travels through hills, tanks and eventually the fields, and has an elevated level of hardness of water, along with pesticide residues, toxins and heavy metals.

3.2.10. Smoking

There was a significantly increased risk of ESRD in renal disease when compared to nonsmokers, in Multiple Risk Factor Intervention Trial. This difference was independent of other factors such as age, gender, diabetes mellitus hypertension, ethnicity etc. [31]. In diabetic individuals, smoking is known to increase the risk of nephropathy. Smoking was identified as a known risk factor for microalbuminuria. Microalbuminuria is recognized to be a marker of renal damage and hence CKD. There is a five times more likely chance of microalbuminuria in the subjects who have a history of smoking when compared to subjects with no history of smoking.

The findings of the study postulate that groundwater, along with cigarette consumption are probable sources of environmental toxin that may lead to CKDu [25].

3.2.11. Cyanobacterial toxin

All the regions that are known to have a high prevalence of the CKD, is clustered around tanks and reservoirs of the irrigation system. A lower prevalence was observed in the communities that used water from natural springs for consumption. The water sampled from these natural springs, showed an absence of cyanobacteria and algae. The water contained very low levels of fluoride, nitrogen, phosphate and potassium. The analysis of the water from the reservoirs showed the presence of cyanobacterial blooms that are capable of producing hepatotoxic effects and carcinogenic effects. The regions that are endemic to CKD are seen to have many reservoirs which could explain the high prevalence of the disease (Athuraliya et al., 2009).

3.2.12. Drinking well water

One study found that in male farmers in PadaviSripura, who spray glyphosate have an increased risk of CKDu coupled with consumption of well water and have a history of consumption from an abandoned well for a period of more than 5 years. Univariable analysis found that consumption of water from an abandoned well increased the risk of contracting CKDu by 7 times. The abandonment of wells that were previously used for drinking water is a commonly observed phenomenon in CKDu endemic areas. The reasoning provided by the locals is that the water is not palatable and therefore unsuitable for drinking or cooking purposes, due to the increase in hardness of water. This leads to the locals having to travel further to obtain water, however more and more wells are being abandoned due to the progression of hard water. Therefore drinking water from a well and from abandoned wells, are used as significant predictors of CKDu [22].

3.2.13. Dehydration

Due to the increasing hardness of water in wells of the PadaviSripura area, the locals need to travel further to obtain water for consumption. This scarcity of water, may lead to less consumption of water leading to dehydration. Dehydration further causes complications to the kidneys, resulting in renal tissue damage, which invariably leads to CKD.

There are more suggested hypotheses put forward in relation to the etiology of CKDu in Sri Lanka. This includes maternal malnutrition, consumption of untreated water from the fields, low intake of calcium, ionic species in water, oxidative stress on renal tissue due to multifactorial origin, genetic predisposition, low BMI, chronic undernutrition, organophosphates, synergy of As and CD, used of NSAIDs, the use of synthetic fertilizers, history of snake bites, consumption of water directly from reservoirs, fungal and bacterial toxins [9].

## IV. CONCLUSION

The Ministry of Health and the World Health Organization estimate that around 15% of the population found in endemic areas are susceptible to develop CKDu. This greatly increases the national socioeconomic burden on the public healthcare system. The specific etiologic causative agent behind CKDu is still unknown. However the suggested hypotheses discussed in this review provides insight for any further research to build on.

## V. ACKNOWLEDGEMENTS

The authors wish to thank the International College of Business and Technology, Sri Lanka for the financial support.